# Cellular automaton model simulating spatiotemporal patterns, phase transitions and evolution concavity in traffic flow

Junfang Tian[1]*, Rui Jiang[2], Guangyu Li[1]*, Martin Treiber[3], Ning Jia[1], Shoufeng Ma[1]

[1]*Institute of Systems Engineering, College of Management and Economics, Tianjin University, Tianjin 300072, China*

[2]*MOE Key Laboratory for Urban Transportation Complex Systems Theory and Technology, Beijing Jiaotong University, Beijing 100044, China*

[3]*Technische Universität Dresden, Institute for Transport & Economics, Würzburger Str. 35, D-01062 Dresden, Germany*

This paper firstly show that a recent model (Tian et al., Transpn. Res. B 71, 138-157, 2015) is not able to well replicate the evolution concavity in traffic flow, i.e. the standard deviation of vehicles increases in a concave/linear way along the platoon. Then we propose an improved model by introducing the safe speed, the logistic function of the randomization probability, and small randomization deceleration for low-speed vehicles into the model. Simulations show that the improved model can well reproduce the metastable states, the spatiotemporal patterns, the phase transition behaviors of traffic flow, and the evolution concavity of traffic oscillations. Validating results show that the empirical time series of traffic speed obtained from Floating Car Data can be well simulated as well.

## 1. Introduction

The formation and evolution of traffic congestion has been investigated for decades (see e.g., Brackstone and McDonald, 1999; Chowdhury, 2000; Helbing, 2001; Nagatani, 2002; Treiber and Kesting, 2013; Kerner, 2004, 2009, 2013; Saifuzzaman and Zheng, 2014; Zheng, 2014). Various traffic flow data have been collected to clarify the nature of traffic flow dynamics (Ranjitkar et al., 2003; Kerner, 2004; Bertini and Monica, 2005; NGSIM, 2006; Laval and Daganzo, 2006; Ahn and Cassidy, 2007; Wagner, 2006, 2010, 2012; Sugiyama et al., 2008; Zheng et al. 2011; Shott, 2011; Tadaki et al., 2013; Laval et al., 2014; Jiang et al. 2014, 2015).

Based on the empirical observations, traditionally two traffic phases are classified: free flow (F) and congested flow. However, by analyzing the long term detector data available to him, Kerner (2004) further distinguished the congested traffic into wide moving jams (J) and synchronized traffic flow (S). The wide moving jam is a moving structure in which vehicles stop or nearly stop. The downstream front of a jam moves upstream with a characteristic speed about 15$km/h$. In contrast, synchronized flow is usually fixed at traffic bottlenecks and the flow rate in the synchronized flow is much larger than that in jam. For the synchronized flow that moves upstream, the propagating speed of its downstream front is not necessarily the characteristic speed.

Kerner claimed that traffic breakdown corresponds to a phase transition from free flow to synchronized flow. Wide moving jams usually emerge from the synchronized flow. As a result, the transition from free flow to jams corresponds to a F→S→J process. In an open road section with an isolated bottleneck, two types of spatiotemporal traffic patterns can be observed: the general pattern and the synchronized flow pattern. In the synchronized flow pattern, which includes the widening synchronized pattern (WSP), the localized synchronized pattern (LSP), and the moving synchronized pattern (MSP), no traffic jam occurs. In contrast, in the general pattern, traffic jam emerges spontaneously from the synchronized flow due to the pinch effect.

---

*Corresponding author.
Email address: jftian@tju.edu.cn, liguangyuli.student@sina.com.

The spontaneous formation of jams is an important feature in traffic flow. Treiterer and Myers (1974) have shown the trajectories of phantom jam, i.e. jams that arise in the absence of any bottleneck. Coifman (1997) also presented the trajectories of thirteen shockwaves in platoons, where small disturbance grows as it propagates upstream until vehicles come to a stop. Sugiyama (2008) and Tadaki et al. (2013) performed traffic experiment on a circuit to investigate the emergence of a jam without bottleneck. The experiments showed that when the density is below a critical value, traffic flow is stable and no jam emerges. However, above another critical density, traffic flow becomes unstable and jam always occurs spontaneously. Between the two critical densities, traffic flow is metestable and the spontaneous formation of jam can be observed probabilistically.

Another important feature in traffic flow is growth process of oscillations. Recently Jiang et al. (2014, 2015) carried out the car following experiments concerning a platoon of 25 passenger cars on a 3.2-km-long open road section. The leading vehicle was asked to move with constant speed. The formation and development of oscillations have been observed. It has been found that standard deviations of velocities increase in a concave or linear way along the 25-car-platoon. For the latter case, it can be expected that if we had a much longer platoon, the variations of speed of cars in the tail of the platoon would be capped due to the physical limits of speeds. Thus the growth rate would bend downward, making the overall curve concave shaped.

In order to simulate the features of traffic flow, many traffic flow models have been proposed. Recently, Tian et al. (2015) have proposed a cellular automaton model, in which there does not exist hypothetical congested steady state, and is therefore named as Non-Hypothetical congested steady state Model (NHM). In NHM, the space gap between two vehicles will oscillate around the (speed dependent) desired value rather than maintaining this value in the deterministic limit in congested traffic flow. This mimics the empirically observed phenomenon that even if the stimulus vanish, the vehicles incline to accelerate or decelerate most of the time, i.e. NHM assumes that there are no congested steady states at all, whether hypothetical or not. It was shown that the model is able to simulate the spatiotemporal patterns and the phase transition behavior in traffic flow.

Nevertheless, as shown in Section 2.1, the NHM has two deficiencies: (i) On the one hand, the NHM is not able to well replicate the evolution concavity in traffic flow; (ii) On the other hand, the cell length is set to $7.5m$. As a result, the model cannot describe the detailed structure of traffic flow. Motivated by this fact, this paper proposes an improved NHM, which is able to simultaneously well reproduce the spatiotemporal patterns, the phase transition behavior as well as the evolution concavity in traffic flow.

The paper is organized as follows. Section 2.1 reviews the original NHM and shows its deficiency. Section 2.2 analyzes the revised NHM with the refined cell length $0.5m$. Section 2.3 presents the improved NHM. Section 3 carries out simulations with the improved model and shows that it can well reproduce the empirical and experimental findings. Section 4 tests the performance of the improved model by the Floating Car Data. Section 5 makes a discussion on state-of-the-art of traffic flow modeling. Section 6 concludes the paper.

## 2. The improved NHM

*2.1 NHM*

For self-consistency of the paper, let us briefly review the NHM. The update rules of NHM are:
1. Acceleration:

$v_n(t+1)=\min(v_n(t)+a, v_{\max})$.
2. Deceleration:
$v_n(t+1)=\min(v_n(t+1), d_{n,\text{eff}}(t))$.
3. Randomization with probability $p_n(t+1)$:
if $(\text{rand}()<p_n(t+1))$ then $v_n(t+1)=\max(v_n(t+1)-b_{\text{rand}}(t+1), 0)$.
4. Determination of $t_{n,\text{st}}(t+1)$:
if $(v_n(t+1)=0)$ then $t_{n,\text{st}}(t+1)=t_{n,\text{st}}(t)+1$,
if $(v_n(t+1)>0)$ then $t_{n,\text{st}}(t+1)=0$.
5. Car motion:
$x_n(t+1)=v_n(t+1)+x_n(t)$.

Here $d_{n,\text{eff}}(t)=d_n(t)+\max(v_{\text{anti}}(t)-g_{\text{safety}}, 0)$ is the effective gap and $v_{\text{anti}}(t)=\min(d_{n+1}(t), v_{n+1}(t)+a, v_{\max})$ is the expected speed of the preceding vehicle in the next time step. $g_{\text{safety}}$ is a safety parameter to avoid accidents and the constraint $g_{\text{safety}} \geq b_{\text{rand}}$ should be satisfied. $a$ is the average acceleration. The randomization probability $p_n(t+1)$ and randomization deceleration $b_{\text{rand}}$ are defined as:

$$p_n(t+1)=p_a H(v_n(t)T-d_{n,\text{eff}}(t))+p_b \Delta(v_n(t),0)H(t_{n,\text{st}}(t)-t_c)+p_c \qquad (1)$$
$$b_{\text{rand}}(t+1)= b_{\text{defens}} H(v_n(t)T-d_{n,\text{eff}}(t))+a \qquad (2)$$

where $T$ is the effective safe time gap. The Heaviside's function $H(x)=1$ if $x>0$, otherwise $H(x, y)=0$. The function $\Delta(x, y)=1$ if $x=y$, otherwise $\Delta(x, y)=0$. $t_{n,\text{st}}(t)$ denotes the time since the last stop for standing vehicles, while $t_{n,\text{st}}(t)=0$ for moving vehicles. NHM can reproduce the synchronized traffic flow.

**Table 1.** Parameter values of NHM.

| Parameters | $L_{\text{cell}}$ | $L_{\text{veh}}$ | $v_{\max}$ | $T$ | $p_a$ | $p_b$ | $p_c$ | $a$ | $b_{\text{defense}}$ | $g_{\text{safety}}$ | $t_c$ |
|---|---|---|---|---|---|---|---|---|---|---|---|
| Units | $m$ | $L_{\text{cell}}$ | $L_{\text{cell}}/s$ | $s$ | - | - | - | $L_{\text{cell}}/s^2$ | $L_{\text{cell}}/s^2$ | $L_{\text{cell}}$ | $s$ |
| Value | 7.5 | 1 | 5 | 1.8 | 0.95 | 0.55 | 0.1 | 1 | 1 | 2 | 8 |

The parameter values of NHM are shown in Table 1. Fig.1 is the simulation results of Jiang's car following experiment (see Section 3.3 for the detail instruction of the simulation setup). The parameters of the model are can be found in Tian et al. (2015). Since the cell length is set to $L_{\text{cell}} = 7.5m$, the NHM cannot describe the detailed structure of traffic flow. Specifically, the velocity of leading car can only equal to $27km/h$ ($1L_{\text{cell}}/s$), $54km/h$ ($2L_{\text{cell}}/s$), $81km/h$ ($3L_{\text{cell}}/s$)... As a result, one can only make a rough comparison between simulation and experiment. Fig.1 shows that even a rough comparison demonstrates that the standard deviation is largely deviated from the experimental data and the spatiotemporal characteristics are not consistent with the experimental one.

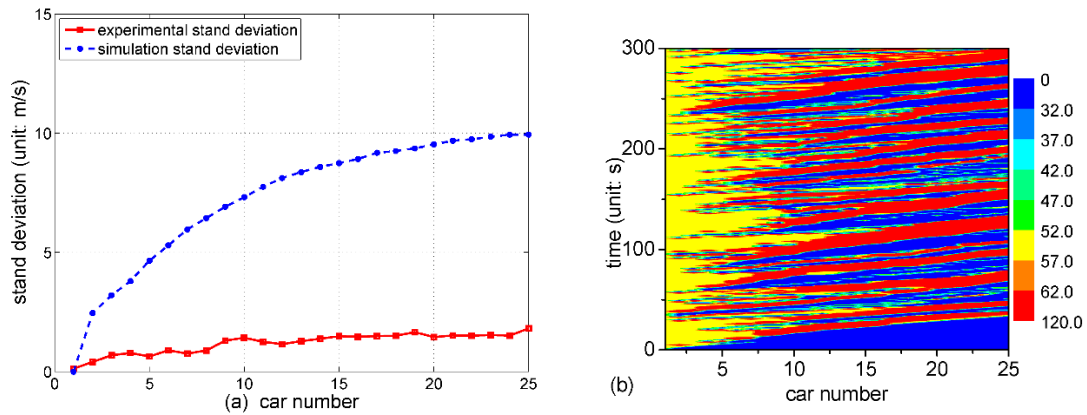

**Fig. 1.** (a) Comparison of the simulation result and experiment result of the standard deviation of the velocities of the cars and (b) the spatiotemporal patterns of the platoon traffic when the leading vehicle moves with the constant speed $v_l=54km/h$. In the experiment, the leading vehicle is required to move with $v_l=50km/h$.

## 2.2 NHM with refined cell length

To better describe the detailed structure of traffic flow, we set the cell length $L_{cell} = 0.5m$, and parameter values are shown in Table 2. Fig.2 shows the flow-density and speed-density diagrams of the model with refined cell length (see Section 3.1 for the detail instruction of the simulation setup). As a comparison, the results of NHM with $L_{cell}=7.5m$ are also presented, which shows that the flow-density and speed-density diagrams are significantly different from that of NHM with $L_{cell}= 0.5m$. The homogeneous initial condition results in the coexistence state of synchronized flow and free flow in NHM with $L_{cell} = 7.5m$ (Fig.3(a)), however, it leads to a strange state in NHM with $L_{cell} = 0.5m$ (Fig.4(a)). As to the megajam initial condition, both models have simulated the coexistence state of wide moving jam and free flow (Fig.3(b) and Fig.4(b)). However, in the density range $K>K_1$, the flux and speed initiated from the homogeneous initial condition are always higher than that from the megajam initial condition, which is also inconsistent with that of NHM with $L_{cell} = 7.5m$ and not in line with the reality since in the high density region, wide moving jams will emerge finally.

**Table 2.** Parameter values of NHM with refined cell length.

| Parameters | $L_{cell}$ | $L_{veh}$ | $v_{max}$ | $T$ | $p_a$ | $p_b$ | $p_c$ | $a$ | $b_{defense}$ | $g_{safety}$ | $t_c$ |
|---|---|---|---|---|---|---|---|---|---|---|---|
| Units | $m$ | $L_{cell}$ | $L_{cell}/s$ | $s$ | - | - | - | $L_{cell}/s^2$ | $L_{cell}/s^2$ | $L_{cell}$ | $s$ |
| Value | 0.5 | 15 | 75 | 1.8 | 0.95 | 0.55 | 0.1 | 1 | 1 | 2 | 8 |

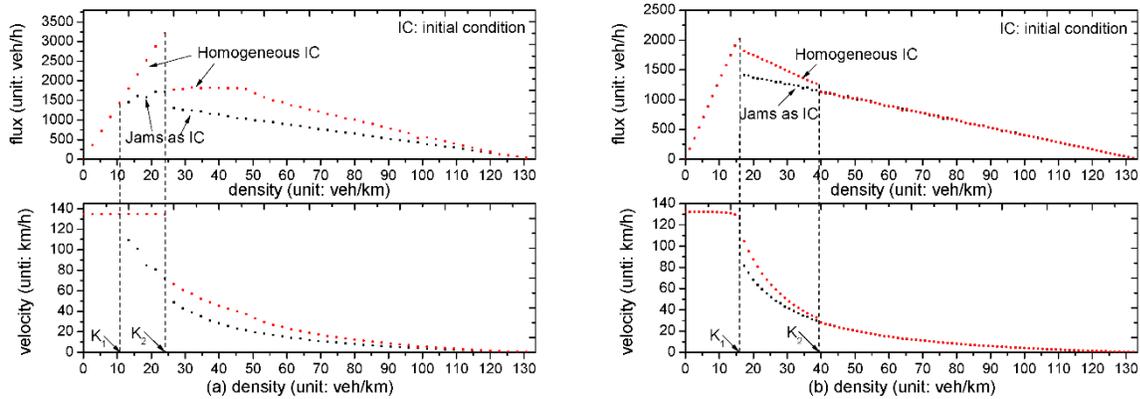

**Fig.2.** Flow-density and speed-density diagrams of (a) NHM with $L_{cell} = 0.5m$ and (b) NHM with default $L_{cell} = 7.5m$. "Homogeneous IC" represents the initially homogenous distribution of traffic, and "Jams as IC" represents the initially megajam distribution of traffic.

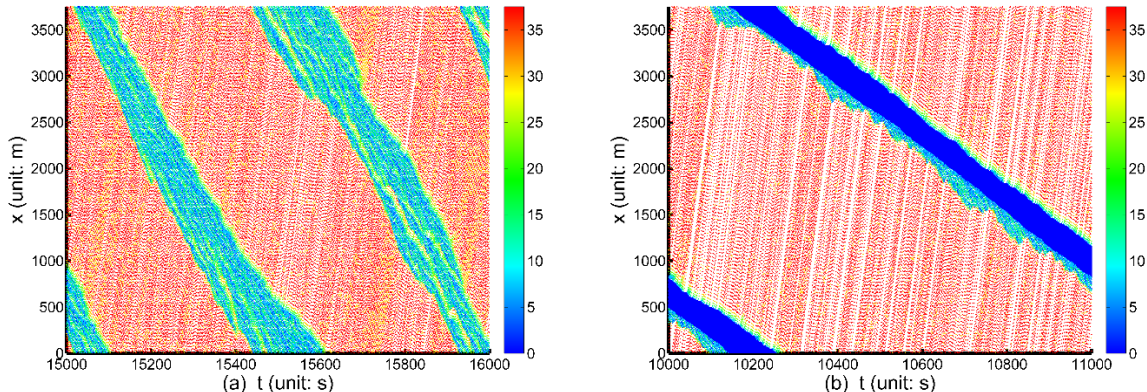

**Fig.3.** The spatiotemporal diagrams of NHM with $L_{cell} = 7.5m$ on the circular road with the density $k=24veh/km$. In (a) the traffic starts from a homogenous initial distribution. In (b) the traffic starts from a megajam. The horizontal direction (from left to right) is time and the vertical direction (from down to up) is space.

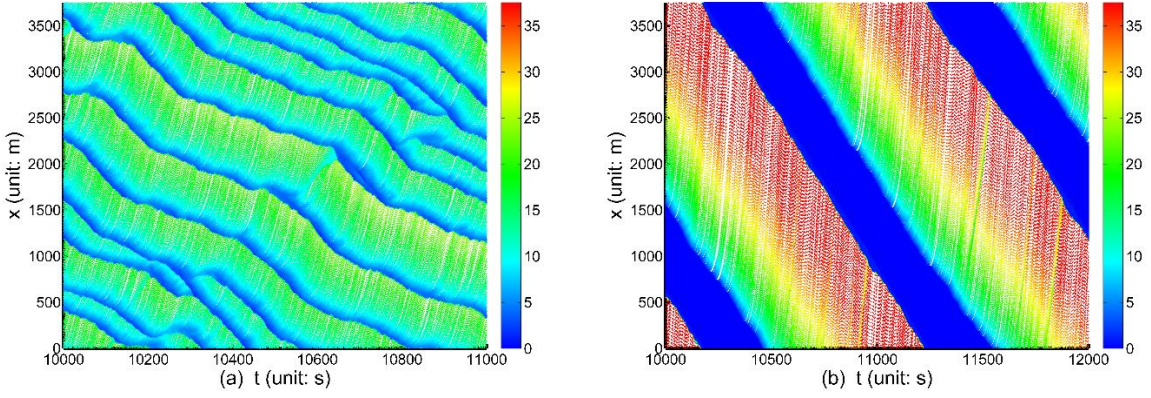

**Fig.4.** The spatiotemporal diagrams of NHM with $L_{\text{cell}} = 0.5m$ on the circular road with the density $k=48 veh/km$. In (a) the traffic starts from a homogenous initial distribution. In (b) the traffic starts from a megajam.

*2.3. The improved NHM*

In the NHM, there exist intrinsic large fluctuations. We introduce the safe speed, the logistic function for the randomization probability and small randomization deceleration for low-speed vehicles into the NHM to avoid large decelerations and large fluctuations. Small randomization probability and randomization deceleration for low-speed vehicles presume that the average decelerations of low-speed vehicles are smaller than that of high-speed vehicles, which is supported by the experimental data shown in Table 3. One can see that the mean deceleration of all following vehicles in Jiang's car-following experiments in low velocity ($7km/h$) is much smaller than that of vehicles in high speed states (no less than $15km/h$).

**Table 3.** Mean decelerations of all following vehicles in Jiang's car-following experiments (Jiang et al. (2014)).

| $v_l$ (unit: $km/h$) | 7 | 15 | 30 | 40 | 50 |
|---|---|---|---|---|---|
| mean deceleration (unit: $m/s^2$) | -0.111 | -0.278 | -0.221 | -0.239 | -0.25 |

Therefore, the deceleration step is revised as:
2. Deceleration:
   $v_n(t+1)=\min(v_n(t+1), d_{n,\text{eff}}(t), v_{n,\text{safe}}(t))$.

$v_{n,\text{safe}}(t)$ is the safe speed defined by the simplified Gipps model (see Equation 11.10 of Treiber and Kesting, 2013):

$$v_{n,\text{safe}}(t) = \text{round}(-b_{\max} + \sqrt{b_{\max}^2 + v_{n+1}(t)^2 + 2b_{\max}d_n(t)}) \qquad (3)$$

This safe speed is applied to keep safety if the leading vehicle $n+1$ decelerates with the maximum deceleration $b_{\max}$ to a complete stop. The function round() returns the value of a number rounded to the nearest integer. The randomization probability $p_n(t+1)$ and randomization deceleration $b_{\text{rand}}$ are redefined as:

$$p_n(t+1) = \frac{p_a}{1+e^{\alpha(v_{c1}-v_n(t))}} H(v_n(t)T - d_{n,\text{eff}}(t)) + p_b \Delta(v_n(t),0) + p_c \qquad (4)$$

$$b_{\text{rand}}(t+1) = b_{\text{defens}} H(v_n(t)T - d_{n,\text{eff}}(t)) H(v_n(t) - v_{c2}) + a \qquad (5)$$

where $\alpha$ and $v_{c1}$ are the steepness and midpoint of the logistic function. Since the fluctuations are greatly diminished and velocity in the synchronized will not decrease to zero, step 4 (i.e. Determination of $t_{n,\text{st}}(t+1)$) can be removed.

In the improved model, the steady states of congested traffic still do not exist. The space gaps oscillate around the desired gap, which is fulfilled by the rule that drivers tend to keep the effective gap no smaller than $v_n(t)T$, otherwise drivers will become defensive by increasing the spontaneous braking probability from $p_c$ to $p_a+p_c$ and the associated deceleration from 1 to $b_{\text{rand}}$. Since the safe speed is introduced, this model is named as the Non-Hypothetical congested steady state with Safe speed cellular automaton Model (NHSM).

## 3. Simulation analysis

Firstly, traffic flow on a circular road and on an open road with an on-ramp are reproduced to simulate the findings of Sugiyama et al. (2008) and Tadaki et al. (2013), and Kerner (2004, 2009), respectively. Then we simulate the car following experiments shown by Jiang et al. (2014, 2015).

During the simulation, the cell length and vehicle length are set as $0.5m$ and $7.5m$, respectively, i.e. $L_{cell} = 0.5m$ and $L_{veh} = 15L_{cell} = 7.5m$. One time step corresponds to $1s$. The parameter values of NHSM are shown in Table 4.

**Table 4.** Parameter values of NHSM.

| Parameters | $L_{cell}$ | $L_{veh}$ | $v_{max}$ | $T$ | $p_a$ | $p_b$ | $p_c$ | $a$ | $b_{max}$ | $b_{defense}$ | $g_{safety}$ | $v_{c1}$ | $v_{c2}$ | $\alpha$ |
|---|---|---|---|---|---|---|---|---|---|---|---|---|---|---|
| Units | $m$ | $L_{cell}$ | $L_{cell}/s$ | $s$ | - | - | - | $L_{cell}/s^2$ | $L_{cell}/s^2$ | $L_{cell}/s^2$ | $L_{cell}$ | $L_{cell}/s$ | $L_{cell}/s$ | $L_{cell}/s$ |
| Value | 0.5 | 15 | 60 | 1.8 | 0.88 | 0.5 | 0.08 | 1 | 6 | 1 | 20 | 30 | 5 | 10 |

*3.1 Circular road*

The following two initial configurations are used in the simulations: 1) all vehicles are homogeneously distributed on the road; 2) all vehicles are distributed in a megajam. Fig.5 shows the flow-density and velocity-density diagrams of NHSM, where the upper two branches ($K_1 < K < K_2$ and $K_2 < K < K_3$) are from the initial homogeneous distribution, while the lower branch is from the initial megajam ($K_1 < K < K_3$). In the branch that the density is smaller than $K_1$, there is only free flow on the road (Fig.6(a)). In the upper branch when the density is in the range of $K_1 < K < K_2$, the synchronized flow and free flow coexist (Fig.6(b)) and the speed in synchronized flow is very high; while in the upper branch that the density belongs to the range of $K_2 < K < K_3$, the synchronized flow and free flow still can coexist (Fig.6(c)) but the speed in synchronized flow is smaller than that of the density range $K_1 < K < K_2$; with the increase of the density, free flow will disappear and only synchronized flow can be found (Fig.6(d)). When the density is greater than $K_3$, the synchronized flow is unstable, and wide moving jams will appear finally (Fig.6(e)). For the lower branch formed from the initial megajam, traffic will evolve to the state that wide moving jams and free flow coexist (Fig.6(f)). Furthermore, it can be found that when the density is smaller than $K_2$, wide moving jams cannot exist, which means that the synchronized flow is very stable in the range of $K_1 < K < K_2$, in which the speed is greater than $65km/h$. This is different from NHM in which the synchronized traffic flow is always in the metastable stable state or unstable state.

Fig.7 shows the spontaneous transitions from free flow to synchronized flow (F→S) and from synchronized flow to wide moving jams (S→J). Therefore, three traffic phases, the metastable states and two first order transitions (F→S and S→J transitions) are clearly distinguished.

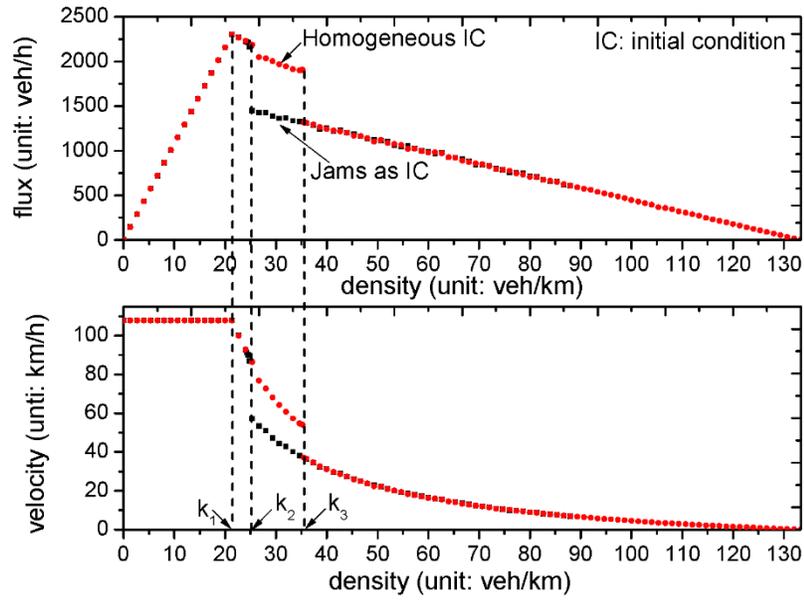

**Fig. 5.** Flow-density and speed-density diagrams of NHSM. "Homogeneous IC" represents the initially homogenous distribution of traffic, and "Jams as IC" represents the initially wide moving jam distribution of traffic.

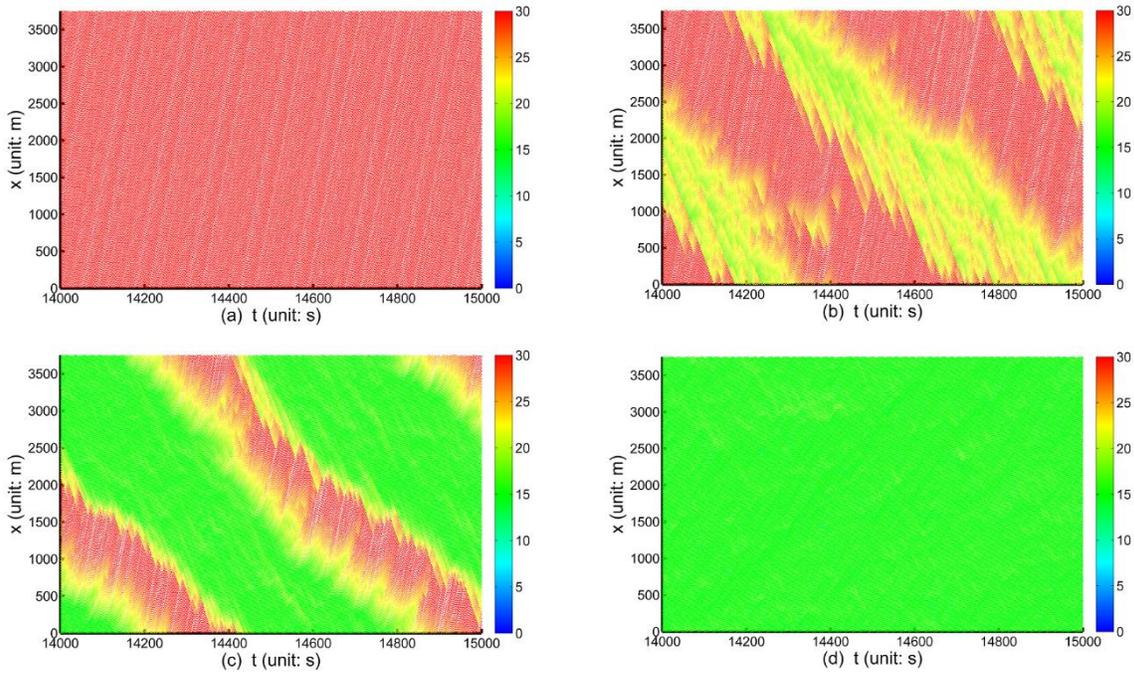

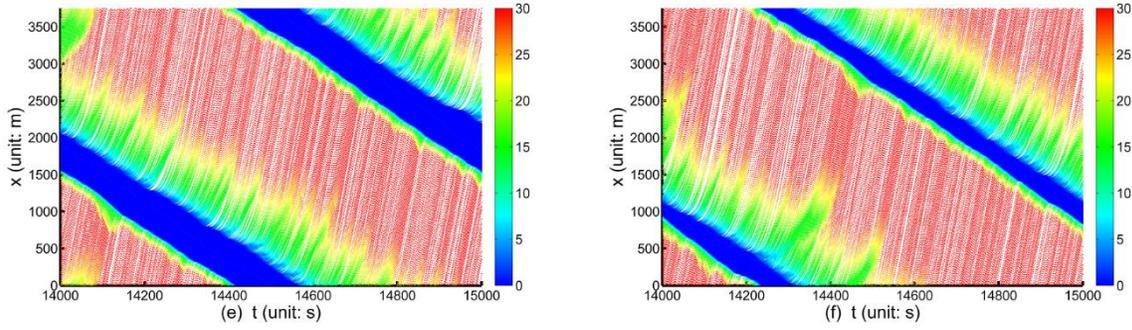

**Fig. 6.** The spatiotemporal diagrams of NHSM on the circular road. In (a-f) the density *k*=21, 24, 29, 35, 29, 37*veh/km*, respectively. In (a-d, f) the traffic starts from a homogenous initial distribution. In (e) the traffic starts from a megajam.

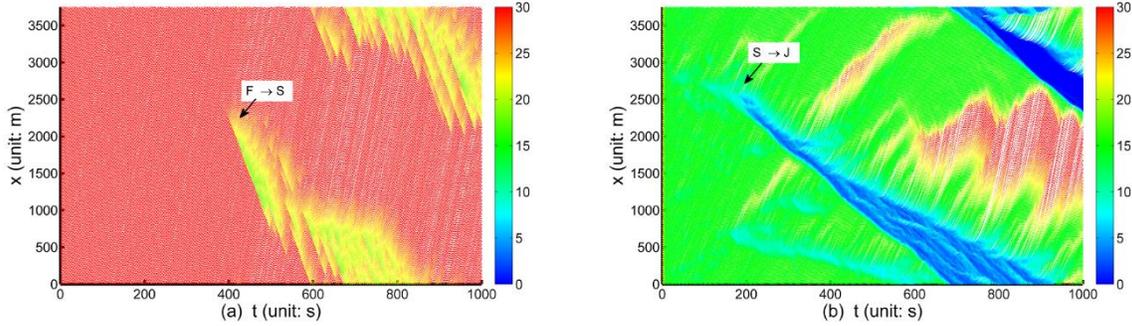

**Fig. 7.** The F→S and S→J transitions of NHSM on the circular road from a homogenous initial distribution. (a) the density *k* = 23*veh/km*; (b) *k* = 40*veh/km*.

*3.2 Open road with an on-ramp*

We simulate the NHSM to study the traffic patterns that emerge near an on-ramp on an open road. $q_{in}$ is the traffic flow entering the main road in units of vehicles per hour and $q_{on}$ is the traffic flow from the on-ramp. When the space gap between two vehicles on the main road in the on-ramp region exceed the value $d_{cri}$ (=$\beta L_{veh} + \gamma v_l$, where $v_l$ is the velocity of the leading vehicle, $\gamma$ = 0.65 and $\beta$ =1.8), a vehicle will be inserted into the middle of them. The length of the on-ramp region is set as $L_{ramp}$=30$L_{veh}$. The details of the simulation setup can be found in Tian et al. (2015).

Fig.8(a) shows the spatiotemporal features of the moving synchronized flow (MSP) where both the upstream and downstream fronts propagate along the upstream direction of traffic flow. Fig.8(b) exhibits the widening synchronized flow (WSP), where only the upstream front propagates upstream and no wide moving jams appear. Fig.8(c) gives the local synchronized pattern (LSP) where the synchronized flow is only localized in the vicinity of the on-ramp region. Fig.8(d) shows the General Pattern (GP) that wide moving jams continuously emerge in the synchronized traffic flow. Therefore, the main empirical findings are successfully simulated by NHSM.

Moreover, the propagation velocity of the downstream of the simulated MSP is about -19*km/h*, which falls into the realistic range between -20 and -10*km/h* (Treiber, et al. 2010). This is a remarkable characteristic of NHSM, since it is better than that in most models that could simulate the synchronized traffic flow, in which the propagation velocity of the boundaries and internal structures of synchronized traffic flow are unrealistic high, for example, in NHM and Kerner-Klenov-Wolf model (KKW, Kerner et al., 2002), the velocity is about -27*km/h* and -30*km/h*, respectively.

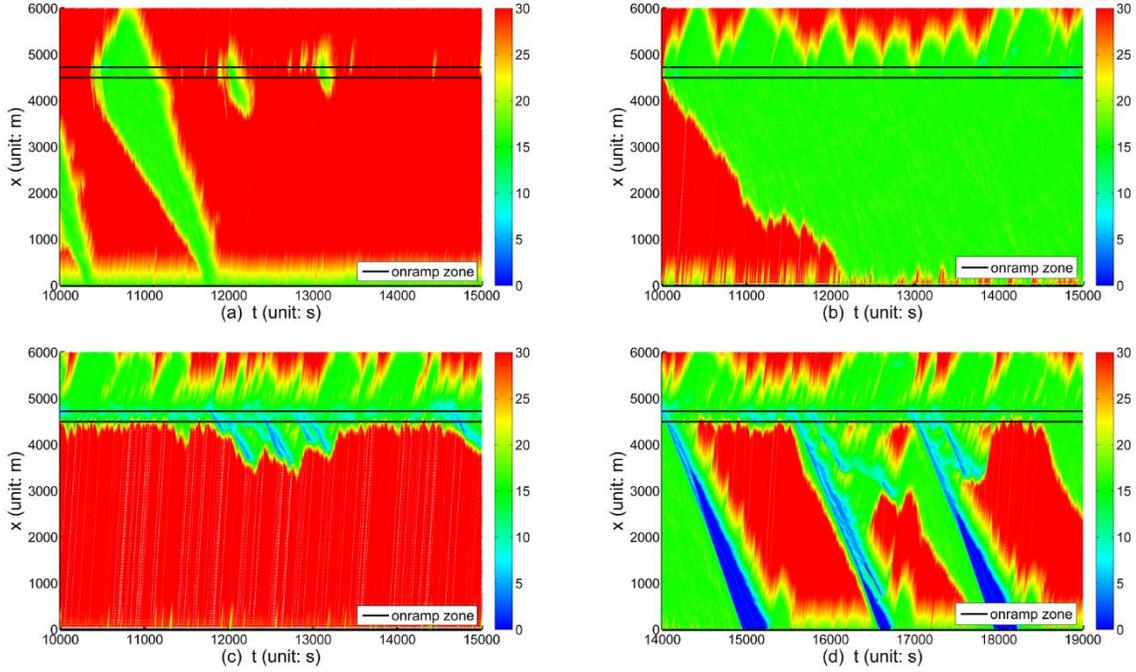

**Fig. 8.** The spatiotemporal diagrams of NHSM on an open road with an on-ramp. (a) $q_{in}$=2042, $q_{on}$ =31 (MSP), (b) $q_{in}$=1957, $q_{on}$ =21 (WSP), (c) $q_{in}$=1623, $q_{on}$ =247 (LSP), (d) $q_{in}$=1812, $q_{on}$ =104 (GP) (unit: veh/h).

*3.3 Car-following experiment simulation*

The platoon of 25 vehicles are investigated. All vehicles are initially distributed in a megajam. The leading vehicle accelerates freely to a prescribed velocity $v_l$ and then moves constantly with $v_l$. The simulation continues for 50000*s* and the first 10000*s* are discarded to let the transients die out. Fig.9 shows that under various velocity $v_l$, the formation and evolution of oscillations of NHSM look very similar to that of Jiang's experimental data. Fig.10 shows that the growth rate of the disturbances is also consistent with the experimental data.

Furthermore, we compare the speed profile of different following vehicles in Jiang's experiments with the predictions of NHSM. Since NHSM is a stochastic model, the probability distribution of the following vehicles was estimated through repeated simulations of the lead vehicle problems (LVPs, Laval et al., 2014). On a LVP, the leading vehicle is asked to move according to the experimental data, while the following vehicle is modeled by NHSM. Fig.11 summarizes all results for experimental and simulated velocity profiles of the following vehicles under different car following experiments. One can see that all experimental profiles are almost falls into the 90%-probability bands, which means that the movements of the following vehicles can be predicted by NHSM.

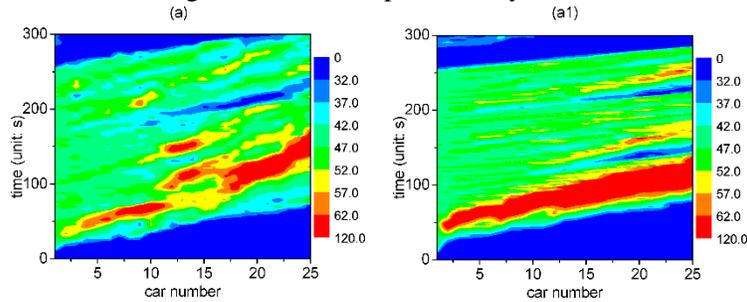

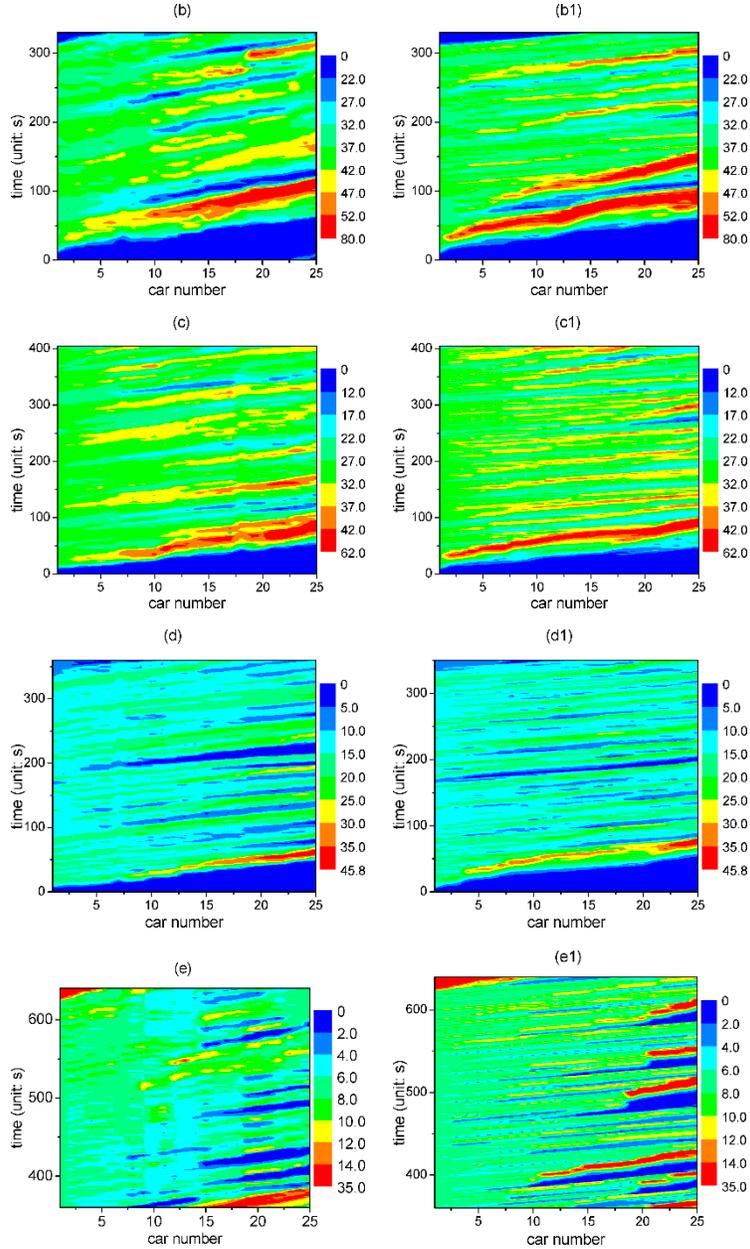

**Fig. 9.** The spatiotemporal patterns of the platoon traffic. The car speed is shown with different colors (unit: *km/h*) as function of time and car number. The left and right Panels show the experimental results and the simulation results of NHSM. In (a, a1) - (e, e1), the leading vehicle is required to move with $v_l$=50, 40, 30, 15, 7*km/h* respectively. In the simulation, the velocity of the leading car is set the same as in the experiment.

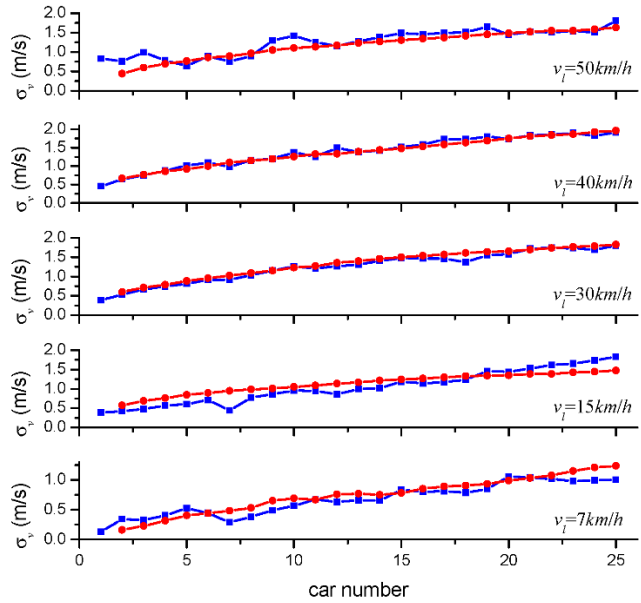

**Fig. 10.** Comparison of the simulation results (symbol solid red lines) and experiment results (symbol solid blue lines) of the standard deviation of the velocities of the cars. The car number 1 is the leading car.

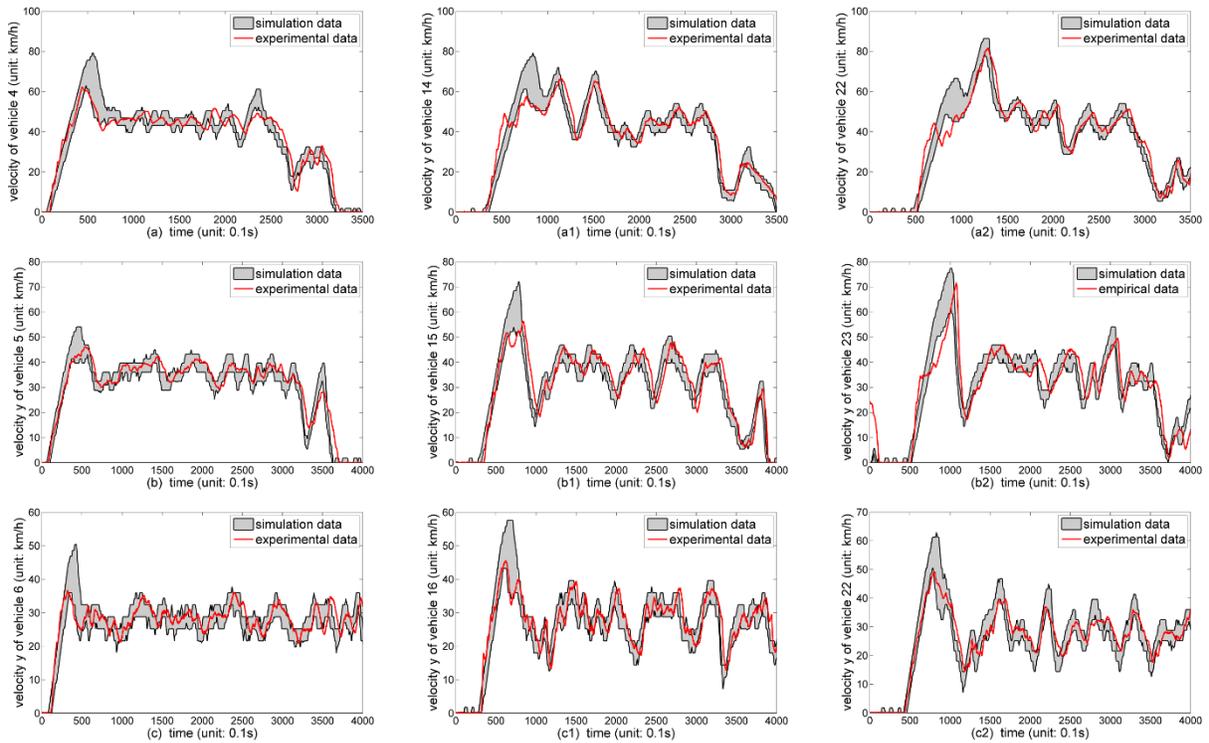

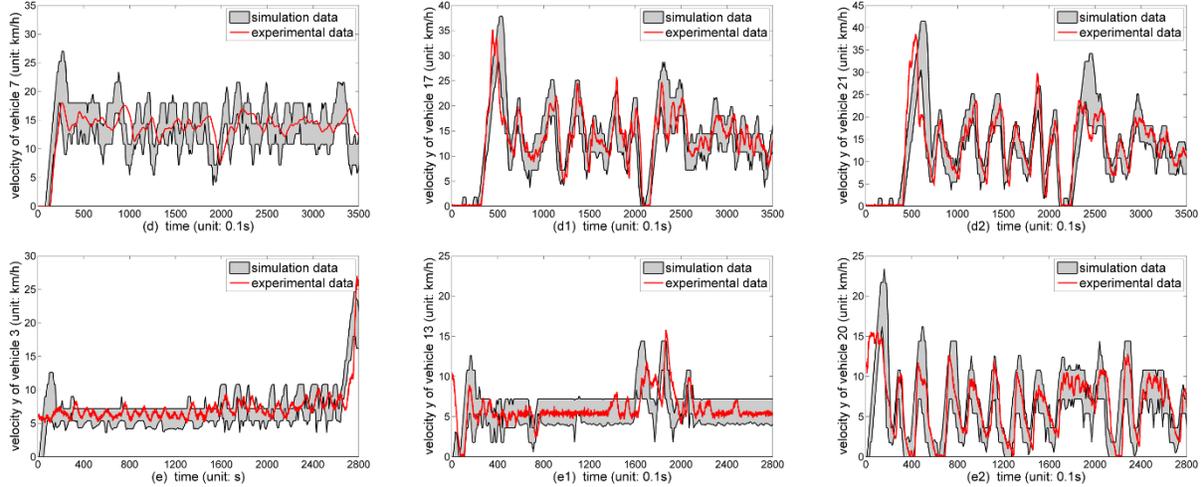

**Fig. 11.** Car following investigations. The red lines are the experimental velocity profiles of different following vehicles taken from Jiang's experiments. The shaded areas are the simulated velocity profiles with 90%-probability band. In (a, a1, a2) - (e, e1, e2), the experimental leading and following vehicles are from different experiments where $v_l$=50, 40, 30, 15, 7$km/h$ respectively.

## 4 Floating Car Data Validation

To test the performance of NHSM, we need to validate it by empirical data sets. Three Floating Car Data (FCD) sets (Kesting and Treiber, 2008) are applied, which are provided by Robert Bosch GmbH (DLR, 2007) and recorded during an afternoon peak period on a fairly straight one-lane road in Stuttgart, Germany. A car equipped with a radar sensor in front provides the relative speed and gap to the car ahead. The durations of the measurements are 400$s$, 250$s$ and 300$s$, respectively. The data are recorded with a frequency of 10 Hz, i.e. with a time increment of 0.1s. The first two sets are limited to the low speed situations below approximately 6$m/s$, corresponding to the gaps smaller than 12$m$ most of the time. In set 3, the speed varies in the range of 0$m/s$ and 18$m/s$, which contains a jump in the gap to the leader from approximately 20$m$ to 40$m$ at a time of about 144s because of a passive lane change of the leader.

During the simulation, we calibrated the model parameters and the results show that the acceleration of NHSM $a$ needs to be adjusted to $2L_{cell}/s^2$ and other parameters are still as shown in Table 2. Fig.12 shows the validation results, which illustrate that MHSM can simulate the FCD very well.

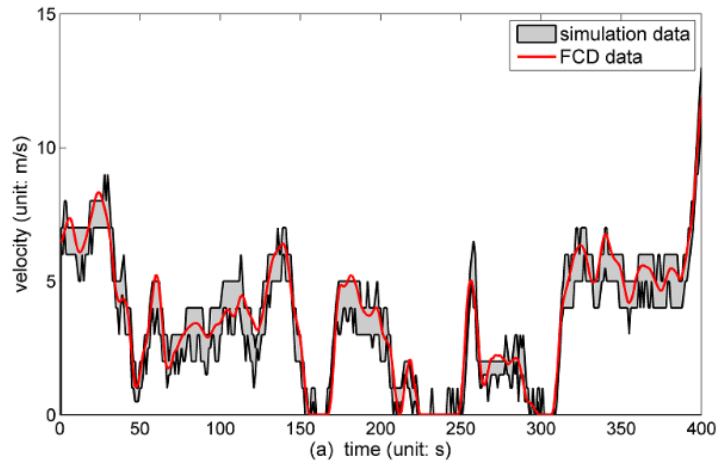

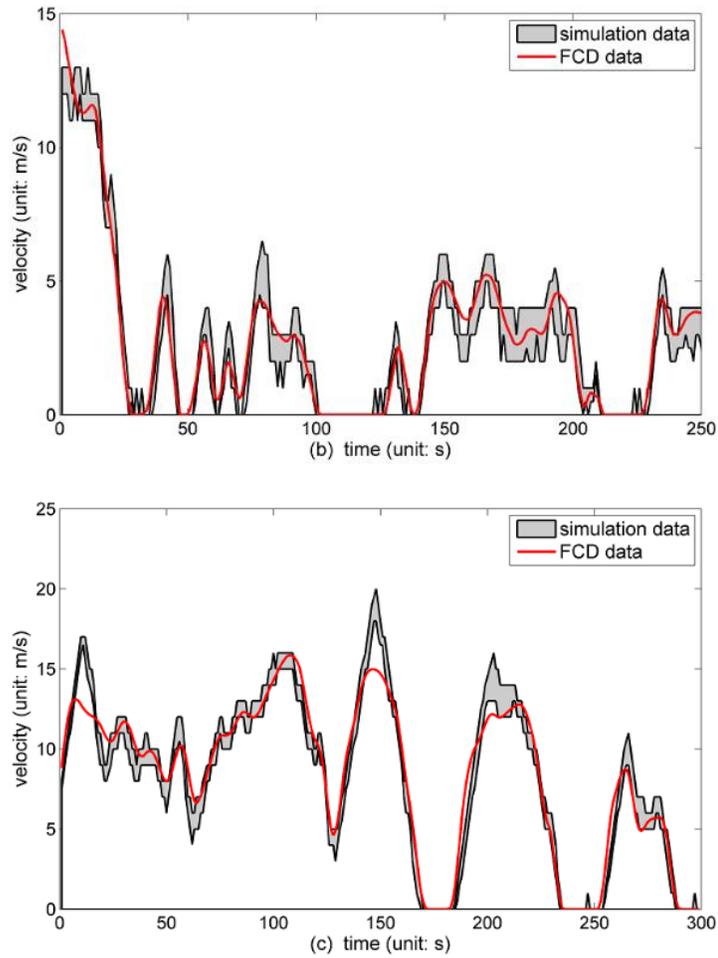

**Fig. 12.** Comparison of simulated and empirical trajectories.

## 5. Discussion

In order to simulate the features of traffic flow, different kinds of models have been proposed, such as the car-following models (Chandler et al. 1958;Newell, 2002;Rakha and Crowther, 2003;Kesting and Treiber, 2008; Laval and Leclercq, 2010;Chen et al. 2012a, 2012b; Aghabayk et al. 2013), the cellular automata models (Nagel and Nagel, 1992;Kerner et al., 2002; Hafstein et al., 2004) and hydrodynamic models (Lighthill and Whitham, 1955; Papageorgiou 1998; Wong and Wong 2002).

In traditional two phase models, such as the General Motor models (GMs, Chandler et al., 1958; Gazis et al. 1961; Edie, 1961), Gipps' Model (Gipps, 1981), Optimal Velocity Model (OVM, Bando et al., 1995), Full Velocity Difference Model (FVDM, Jiang et al., 2001) and Intelligent Driver Model (IDM, Treiber et al., 2000), it is usually assumed either explicitly or implicitly that there is a unique relationship between density and flow in stationary homogeneous traffic of identical vehicles, which is called the fundamental diagram. In the two-phase models, formation of traffic jams is explained via linear instability of the models.

However, Kerner (2004) argued that the two-phase models cannot well reproduce the empirical features of traffic flow, in particular, the traffic breakdown from free flow to synchronized flow. Jiang et al. (2014, 2015) have shown that the simulation results of the two-phase models with additional reasonable acceleration noise run against the experimental finding, in which the standard deviation initially increases in a convex way in the unstable density range.

Kerner proposed the three-phase traffic theory, in which it is hypothesized that the steady states of synchronized flow should cover a two-dimensional region in the flow–density plane. Based on Kerner's three-phase traffic theory, many models have been developed, such as the Kerner-Klenov-Wolf model considering the speed adaptation effect (KKW, Kerner et al., 2002), the model considering mechanical restriction versus human overreaction (Lee et al., 2004), the model considering the brake light effect (Jiang and Wu, 2003, 2005), the model assuming that randomization depends on speed difference (Gao et al. 2007, 2009) and so on. These models have been claimed to be able to simulate the spatiotemporal patterns of traffic flow.

Recently, Jiang et al. (2014) also found that by removing the fundamental diagram and allowing the traffic state to span a two-dimensional region in velocity-spacing plane, the growth pattern of disturbances has changed and becomes qualitatively in accordance with the observations. In particular, the 2-Dimensional Intelligent Driver Model (2D-IDM) considering variable desired time headway can fit the experimental results quantitatively well.

Now a question arises naturally, i.e., can these models simultaneously well simulate the known traffic flow features as the NHSM? We have carried out simulations with several models mentioned above. Unfortunately none of them has been found to perform as well as the NHSM. Specifically, the KKW model, the model of Lee et al., the brake light model, and the model of Gao et al. cannot well reproduce the evolution concavity of the traffic flow. The 2D-IDM fails to simulate the synchronized flow pattern. Nevertheless, on the one hand, the NHSM is cellular automaton model. Although the cell length is set to 0.5$m$, the model is still coarse-grained comparing with car following models which are continuous in both time and space. Therefore, efforts are needed in the field of traffic flow modeling to establish well-performing car-following models. On the other hand, the ability of NHSM to depict traffic features to be revealed by future data needs to be carefully investigated in the future work.

## 6. Conclusion

Traffic flow models play significant role in traffic micro-simulations. Models that can reproduce all identified traffic phenomena are always pursued to flourish and advance traffic flow theory. In a recent paper, Tian et al. (2015) have proposed a NHM model based on the assumption that the space gap between two vehicles will oscillate around the (speed dependent) desired value rather than maintain this value in the deterministic limit in congested traffic flow. Although the NHM has successfully simulated the spatiotemporal patterns of traffic flow, it is not able to well replicate the evolution concavity in traffic flow.

This paper proposes an improved NHM by introducing the safe speed and the logistic function of the randomization probability. Simulations show that the improved NHM can well reproduce all identified phenomena. Specifically, simulations on an open road with a moving bottleneck (realized by a slow moving leading vehicle in a platoon) illustrate that the evolution concavity of traffic oscillations can be well reproduced. Simulations on a circular road show that the improved model is able to depict the metastable state, free flow, synchronized flow, jams as well as the transitions among the three phases. Simulations on an open road with an on-ramp demonstrate that the spatiotemporal patterns of traffic flow can be well described. Validating results show that the empirical time series of traffic speed obtained from Floating Car Data can be well simulated.

In the future work, more traffic data (including empirical data and larger-scale experimental data), in particular, traffic data related to traffic breakdown from free flow to synchronized flow, need to be collected. Then the improved NHM as well as other models should be examined by using these data.

**Acknowledgements:**

The authors wish to thank NGSIM for supplying the empirical data used in this article. This work was supported by the National Basic Research Program of China under Grant No. 2012CB725400. JFT was supported by the National Natural Science Foundation of China (Grant No. 71401120). RJ was supported by the Natural Science Foundation of China (Grant Nos. 11422221 and 71371175). NJ was supported by the National Natural Science Foundation of China (Grant

No. 71101102). SFM was supported by the National Natural Science Foundation of China (Grant No. 71431005).No. 71101102). SFM was supported by the National Natural Science Foundation of China (Grant No. 71431005).